\documentclass[]{emulateapj}
\usepackage{graphicx}
\begin{document}

\title{Galaxy Zoo Morphology and Photometric Redshifts in the Sloan Digital
Sky Survey} 

\author{M. J. Way\altaffilmark{1,2}}
\affil{NASA Goddard Institute for Space Studies,
2880 Broadway, New York, NY 10029, USA}
\altaffiltext{1}{NASA Ames Research Center, Space Sciences Division,
MS 245-6, Moffett Field, CA 94035, USA}
\altaffiltext{2}{Department of Astronomy and Space Physics,
Uppsala, Sweden}

\begin{abstract}
It has recently been demonstrated that one can accurately derive galaxy
morphology from particular primary and secondary isophotal shape estimates
in the Sloan Digital Sky Survey imaging catalog. This was accomplished
by applying Machine Learning techniques to the Galaxy Zoo morphology catalog.
Using the broad bandpass photometry of the Sloan Digital Sky Survey in
combination with with precise knowledge of galaxy morphology should help in
estimating more accurate photometric redshifts for galaxies. Using the
Galaxy Zoo separation for spirals and ellipticals in combination with
Sloan Digital Sky Survey photometry we attempt to calculate photometric
redshifts. In the best case we find that the root mean square error for Luminous
Red Galaxies classified as ellipticals is as low as 0.0118. Given these
promising results we believe better photometric redshift estimates for all
galaxies in the Sloan Digital Sky Survey ($\sim$350 million) will be feasible
if researchers can also leverage their derived morphologies via Machine Learning.
These initial results look to be promising for those interested in estimating
Weak-Lensing, Baryonic Acoustic Oscillation, and other fields dependent upon
accurate photometric redshifts.

\end{abstract}

\keywords{galaxies: distances and redshifts --- methods: statistical}


\section{Introduction}\label{sec:intro}

It is commonly believed that adding information about the morphology
of galaxies may help in the estimation of Photometric Redshifts (Photo-Zs)
when using training set methods. Most of this work in recent years has utilized
The Sloan Digital Sky Survey \citep[SDSS,][]{York2000}.  For example, as
discussed in \citet[][hereafter Paper II]{Way09} many groups have attempted to use
a number of derived primary and secondary isophotal shape estimates in the
Sloan Digital Sky Survey imaging catalog to help in estimating Photo-Zs.
Some examples include; using the radius containing
50\% and/or 90\% of the \cite{Petrosian1976} flux in the SDSS r band (denoted as
petroR50\_r petroR90\_r in the SDSS catalog), concentration index
(CI=petroR90\_r/petroR50\_r), surface brightness, axial ratios and
radial profile \citep[e.g.][]{CL2004,Ball2004,Wadadekar05,Kurtz2007,WG08}.

More recently \cite{Singal2011} have attempted to use Galaxy Shape
parameters derived from Hubble Space Telescope/Advanced Camera for Surveys
imaging data using a principle
components approach and then feeding this information into their
Neural Network code to predict Photo-Zs, but for samples
much deeper than the SDSS. Unfortunately they find marginal improvement
when using their morphology estimators.

Another promising approach focuses on the reddening and inclination of
galaxies. \cite{Yip2011} have attempted to quantify these effects on a galaxy's
spectral energy distribution (SED).  The idea is to use this information to
correct the over-estimation of Photo-Zs of disk galaxies.

On the other hand, attempts to morphologically classify large number of galaxies
in the universe has gained in accuracy over the past 15 years as better/larger
training samples
from eye classification has increased. For example, \cite{Lahav1995} was one
of the first to use an Artificial Neural Network trained on 830 galaxies
classified by the eyes of six different professional astronomers.
In more recent years \cite{Ball2004} has attempted to classify galaxies by
morphological type using a Neural Network approach based on a sample of 1399
galaxies (from the catalog of \cite{Nakamura2003}). \cite{Cheng2011} has
used a sample of 984 non-star forming SDSS early-type galaxies to distinguish
between E, S0 and Sa galaxies.  In the past year two
new attempts at morphological classification using Machine Learning techniques
on a Galaxy Zoo \citep{Lintott2008,Lintott2011} training sample have 
been published \citep{Banerji2010,HC2011}.
The \cite{Banerji2010} results were impressive in that they claim to obtain
classification to better than 90\% for three different morphological classes
(spiral, elliptical and point-sources/artifacts).

These works are in contrast to previous work like that of \cite{Bernardi2003}
who used a classification scheme based on SDSS spectra. However, this
classification certainly missed some early-type galaxies from their desired
sample due to the presence of star formation.

In this paper we will continue our use of Gaussian Process Regression to
calculate Photo-Zs, using a variety of inputs.
This method has been discussed extensively in two previous papers
\citep{Way06,Way09}.

We utilize the SDSS Main Galaxy Sample
\citep[MGS,][]{Strauss02} and the Luminous Red Galaxy Sample
\citep[LRG,][]{Eisenstein01} from the SDSS Data Release Seven
\citep[DR7,][]{SDSS07}. We also utilize the Galaxy Zoo 1
survey results \citep[GZ1,][]{Lintott2011}.
The Galaxy Zoo project\footnote{http://www.galaxyzoo.org} \citep{Lintott2008}
contains a total of 900,000 SDSS galaxies with morphological
classifications \citep{Lintott2011}.

While this study does not focus exclusively on the LRG sample, it should be
noted that if it is possible to improve the Photo-Z estimates for
these objects as shown herein it could also improve the estimation of
cosmological parameters
\citep[e.g.][]{BB2005,Padmanabhan2007,Percival2010,Reid2010,Zunckel2011} using the
SDSS as well as upcoming surveys such as
BOSS\footnote{Baryon Oscillation Spectroscopic Survey}\citep{Cuesta2011,Eisenstein2011},
BigBOSS \citep{Schlegel2009}, and possibly Euclid \citep{Sorba2011},
not to mention LSST\footnote{Large Synoptic Survey Telescope}\citep{Ivezic08}.
It could also contribute to more reliable Photo-Z errors, as required for
weak-lensing surveys \citep{Bernstein2010,Kitching2011} and Baryonic Acoustic
Oscillation measurements, which are also dependent upon accurate
Photo-Z estimation of LRGs \citep{Roig2008}.

\section{Data}\label{sec:section1}

All of the data used herein have been obtained via the SDSS casjobs
server\footnote{http://casjobs.sdss.org}. In order to obtain results consistent
with Paper II for both the MGS and LRG samples we use the same photometric
quality flags (!BRIGHT and !BLENDED and !SATURATED) and redshift
quality (zConf$>$0.95 and zWarning=0) but using the SDSS DR7
instead of earlier SDSS releases.  These data are cross-matched in casjobs with
columns 14--16 in Table 2 of \cite{Lintott2011} extracting the galaxies
flagged as `spiral', `elliptical' or `uncertain'. The galaxies ``flagged
as `elliptical' or `spiral' require 80 per cent of the vote in that category
after the debiasing procedure has been applied; all other galaxies are
flagged `uncertain'" \citep{Lintott2011}. Debiasing is the processes of
correcting for small biases in spin direction and color.
See Section 3.1 in \cite{Lintott2011} for more details on debiasing.

Note that the GZ1 sample is based upon the MGS, but the MGS contains
LRGs as well. This is why we can analyze both of these samples. However, the actual
LRG survey goes fainter than the MGS and so we do not find LRG galaxies
fainter than the MGS limit of r$_{petrosian}\lesssim$17.77.
See \cite{Strauss02} and \cite{Eisenstein01} for details on the MGS and LRG samples.

\begin{figure}[htb]
\centering
\includegraphics[scale=0.4]{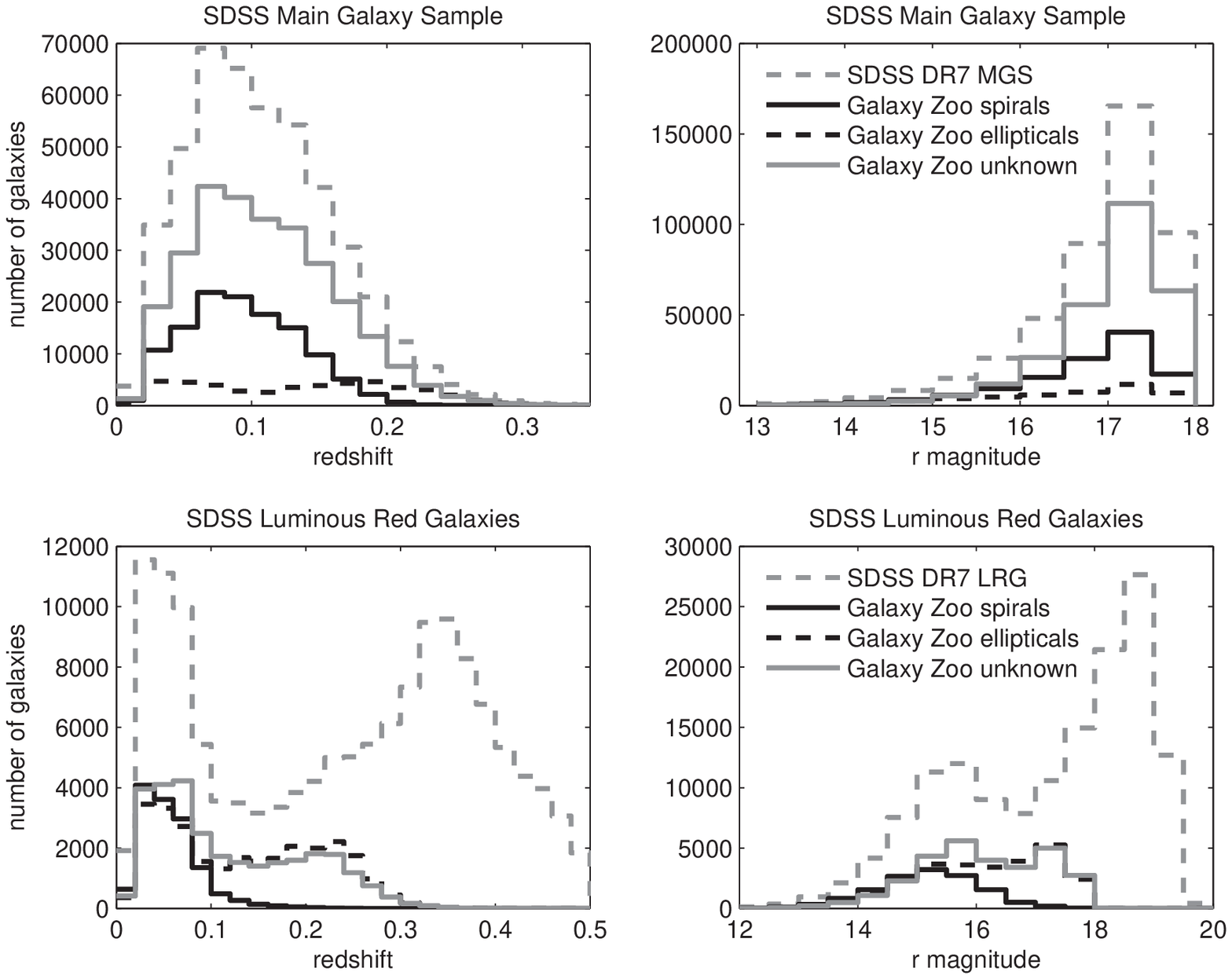}
\caption{Redshift and r-band dereddened model magnitudes for the
Main Galaxy Sample (top two panels) and Luminous Red Galaxies
(bottom two panels).}\label{fig:dr7hist}
\end{figure}

A number of points from both the LRG and MGS were eliminated because of either
bad values (e.g. -9999) or because they were considered outliers from the main
distribution of points. The former offenders included: petroR90\_i (13 points in the
MGS sample, 1 point in the LRG), mE1\_i (43 points, 5 points), petroR90Err\_i
(7177 points, 1262 points), mRrCcErr\_i (22 points, 12 points).
The reason for eliminating bad mE1\_i points is that we use it for
calculating aE\_i from Table 2 of \cite{Banerji2010}. A small number
of outliers were also removed from the MGS sample, but totalled
only 27 points.  No such outlier points were removed in the LRG sample.
This leaves us with a total of 437,273 MGS and 68,996 LRG objects.
Using the GZ1 classifications in the MGS there are 45,249 ellipticals,
119,369 spirals and 272,655 uncertain ($\sim$ 62\%).
For the LRG sample there are 27,227 ellipticals and 13,495 spirals
leaving 28,274 uncertain ($\sim$41\%).

\section{Discussion}\label{sec:discussion}

Using the morphological classifications from the Galaxy Zoo project
first data release \citep{Lintott2011} we attempt to calculate Photo-Zs
for 4 different samples and four combinations of primary and secondary
isophotal shape estimates from the SDSS as seen in Table \ref{tbl-1}.
A larger variety of input combinations
were tried including those in Table 1 of \cite{Banerji2010}.
However, we only report those found with the lowest root mean square
error (rmse) in Table \ref{tbl-1} of this paper.

\begin{deluxetable}{llc}
\tabletypesize{\scriptsize}
\tablecolumns{3}
\tablecaption{Results\label{tbl-1}}
\tablehead{
\colhead{Data\tablenotemark{a}}   &
\colhead{Inputs\tablenotemark{b}} & \colhead{$\sigma_{rmse}$\tablenotemark{c}}}
\startdata
MGS-ELL	&ugriz+Q+U		& 0.01561 0.01532 0.01620\\
-	&ugriz+P50+CI		& 0.01407 0.01400 0.01475\\
-	&ugriz+P50+CI+Q+U	& 0.01641 0.01560 0.01801\\
-	&ugriz+B		& 0.01679 0.01668 0.01683\\
\tableline
MGS-SP	&ugriz+Q+U		& 0.01889 0.01864 0.01913\\
-	&ugriz+P50+CI		& 0.01938 0.01927 0.01947\\
-	&ugriz+P50+CI+Q+U	& 0.01751 0.01747 0.01777\\
-	&ugriz+B		& 0.02092 0.02089 0.02101\\
\tableline
LRG-ELL	&ugriz+Q+U		& 0.01345 0.01291 0.01420\\
-	&ugriz+P50+CI		& 0.01334 0.01278 0.01426\\
-	&ugriz+P50+CI+Q+U	& 0.01584 0.01439 0.01693\\
-	&ugriz+B		& 0.01180 0.01175 0.01184\\
\tableline
LRG-SP	&ugriz+Q+U		& 0.01520 0.01404 0.01910\\
-    	&ugriz+P50+CI		& 0.01514 0.01474 0.01679\\
-	&ugriz+P50+CI+Q+U	& 0.01957 0.01870 0.02285\\
-	&ugriz+B		& 0.01737 0.01728 0.01765\\
\enddata
\tablenotetext{a}{MGS=Main Galaxy Sample \citep{Strauss02}, 
LRG=Luminous Red Galaxies \citep{Eisenstein01}, SP=Classified as spiral
by Galaxy Zoo, ELL=Classified as elliptical by Galaxy Zoo}
\tablenotetext{b}{u-g-r-i-z=5 SDSS dereddened magnitudes,
P50=Petrosian 50\% light radius
in SDSS i band, CI= Concentration Index (P90/P50), Q=Stokes Q value in i band,
U=Stokes U value in i band, B=Inputs from Table 2 of \cite{Banerji2010}=CI,mRrCc\_i,aE\_i,mCr4\_i,texture\_i}
\tablenotetext{c}{We quote the bootstrapped 50\%, 10\% and 90\% confidence
levels as in Paper II for the root mean square error (rmse)}
\end{deluxetable}

\begin{figure}[htb]
\centering
\includegraphics[scale=0.4]{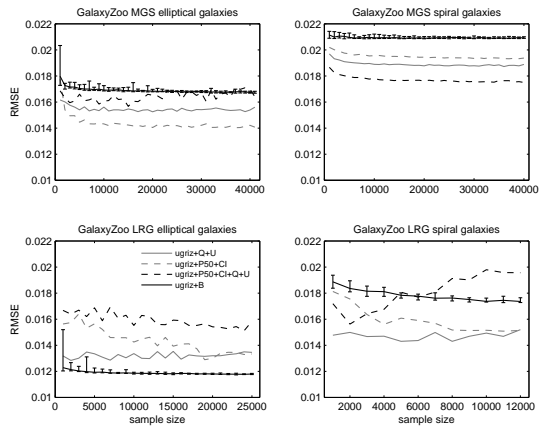}
\caption{Plots of room mean square error for a given number of galaxies per
50\% bootstrap level with representative errors (10\% and 90\%).
Main Galaxy Sample (top two panels elliptical and spiral) and
Luminous Red Galaxies (bottom two panels elliptical and spiral).}
\label{fig:galzoo7PLOT4}
\end{figure}

\begin{figure}[htb]
\centering
\includegraphics[scale=0.4]{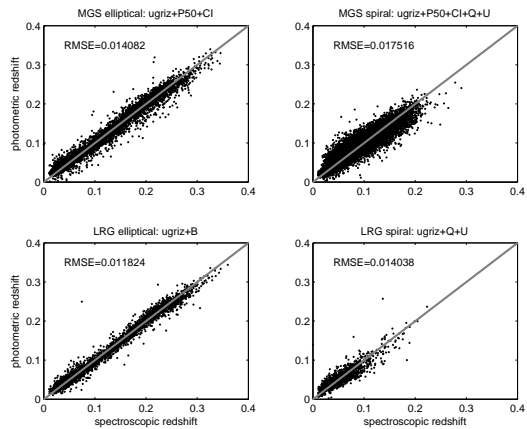}
\caption{Plots of spectroscopic redshift versus predicted photometric
redshift for the input with the lowest rmse for each of the four given data sets
shown in Table \ref{tbl-1}}
\label{fig:galzoo7PLOT4Z}
\end{figure}

The results using the \cite{Banerji2010} suggested isophotal shape estimates
as well as others tested in Paper II are found in Figure \ref{fig:galzoo7PLOT4}
and Table \ref{tbl-1}. In Figure \ref{fig:galzoo7PLOT4Z} we also show plots
of the spectroscopic redshift
versus the predicted photometric redshift for the inputs that predict
the lowest rmse for each of the four data sets listed in Table \ref{tbl-1}.
These are more impressive than one might initially guess. In Paper II we showed
how adding additional bandpasses in the ultraviolet via the Galaxy Evolution
Explorer\footnote{http://www.galex.caltech.edu}\citep[GALEX,][]{Martin05}
could naively improve Photo-Z estimation. The same was shown when using
additional bandpasses from the infrared from the Two Micron All Sky
Survey\footnote{http://www.ipac.caltech.edu/2mass}\citep[2MASS,][]{Skrutskie06}.
However, the results were biased because neither GALEX or 2MASS reach the same
magnitude or redshift depth as the full SDSS MGS or LRG samples.  It is easier
to get lower rmse estimates of Photo-Z when you have a smaller range of lower
redshifts to fit. For the MGS it is clear from the top two
panels in Figure \ref{fig:dr7hist} that the Galaxy Zoo objects span a similar
range of redshifts and r-band magnitudes. On the other hand the situation
for the Luminous Red Galaxies is not as straightforward. Looking at the bottom
two panels of Figure \ref{fig:dr7hist} the large second bump at a redshift
of z$\sim$0.35 and r$\sim$18 does not exist.
The latter is logical because the Galaxy Zoo catalog was drawn from
the MGS and hence there are no galaxies beyond r$_{petrosian}=$17.77
(see \cite{Petrosian1976} for details on Petrosian magnitudes) according
to their selection criteria \citep{Strauss02}.

Our lowest rmse values come from galaxies categorized as ellipticals
in the Luminous Red Galaxy Sample using the SDSS u-g-r-i-z bandpass filters and
the isophotal shape estimates from Table 2 of \cite{Banerji2010}:
ci, mRrCc\_i, aE\_i, mCr4\_i, texture\_i.
These yield an rmse of only 0.01180, which we believe is the lowest calculated
to date for such a large sample of galaxies measured in the bandpasses of the
SDSS while also retaining a fairly large range of redshifts
(0 $\lesssim z \lesssim$ 0.25) and dereddened 
magnitudes (12 $\lesssim r_{petrosian} \lesssim$ 17.77).

Taking a closer look at the kinds of inputs that improve the results by galaxy
type can be interesting. It is clear from Table \ref{tbl-1} that the Stokes
parameters appear to work better for spiral than elliptical galaxies.
The Stokes parameters measure the axis ratio and position angle of galaxies
as projected on the sky.  In detail they are flux-weighted second moments
of a particular isophote.
\begin{equation} M_{xx}\equiv\langle\frac{x^{2}}{r^{2}}\rangle, \ \ \
M_{yy}\equiv\langle\frac{y^{2}}{r^{2}}\rangle, \ \ \
M_{xy}\equiv\langle\frac{xy}{r^{2}}\rangle
\end{equation}
When the isophotes are self-similar ellipses one finds \citep{Stoughton2002}:
\begin{equation}
Q\equiv M_{xx}-M_{yy}=
\frac{a-b}{a+b}\cos(2\phi), \ \ \
U\equiv M_{xy}=\frac{a-b}{a+b}\sin(2\phi),
\end{equation}

The semi-major and semi-minor axes are a and b while $\phi$ is the position
angle. \cite{masters2010} demonstrates the efficacy of using SDSS derived
axis ratios in characterizing the inclinations of spiral galaxies.
This is seen in Table \ref{tbl-1} where they offer the second
best set of inputs when determining photometric redshift for spirals.
Both Stokes Q \& U parameters also display a larger range of values in the
spirals than in the ellipticals.  The standard deviations in Stokes Q \& U for
spirals are 0.1877 \& 0.1500 while for ellipticals they are 0.0596 \& 0.0459.
Hence they clearly offer more room for possible improvement in the former
than in the latter.

One of the more surprising results is the difference in using the B inputs for
the MGS versus LRG ellipticals. In the latter case these inputs
give the lowest RMSE results, while in the MGS elliptical case they give the worst.
This could be do to the fact that the surface brightness of the LRG
galaxies are more easily modeled by the B inputs than the MGS. The MGS 
ellipticals may still have clumps of star formation that can make the surface 
brightness more difficult to model than the more passive LRG ellipticals.

When comparing the MGS and LRG spirals one stark difference is clear when utilizing
the P50 (Petrosian 50\% light radius in SDSS i band) and
CI (Concentration Index=P90/P50) inputs shown in Table \ref{tbl-1}.
In the MGS spiral case these additional inputs yield
worse fits, whereas they are among the most useful in the LRG spiral case.
This may indicate that MGS spirals are more diverse morphologically than
LRG spirals. The P50 and CI inputs are incapable of helping to model the MGS
spiral diversity and simply add noise rather than signal to the fits.
\cite{masters2010} points out that red spirals (read LRG type) will
``be dominated by inclined dust reddened spirals, and spirals with large
bulges.'' Note that this does not mean that LRG bulge dominated
spirals are necessarily S0 galaxies (which would add to their diversity both
morphologically and spectroscopically).  \cite{Lintott2008,Bamford2009} have both
shown that contamination of S0s into spirals is only about 3\% in the best case
scenario.  So again, perhaps P50 and CI can do a better job of modeling LRG
spirals because they are less diverse than MGS spirals.

There are several outstanding issues with using this approach for studies that
may utilize large samples of SDSS LRG derived Photo-Zs
(e.g. Baryonic Acoustic Oscillations).
The first is that the GZ1 catalog has only been able to classify
($\sim$59\%) of the LRG galaxies as spiral or elliptical. This means that 41\%
of our sample cannot benefit from morphology knowledge when estimating Photo-Zs.
Secondly, the LRGs used herein do not go to the same depth (in redshift or magnitude)
as the full LRG (r$\lesssim$19) catalog since the GZ1 is based on
the MGS (r$\lesssim$17.77).
Note also that the GZ1 morphology estimates get worse as one reaches
the fainter end of the sample \citep{Lintott2008}.
Thirdly, the Machine Learning derived morphologies of
\cite{Banerji2010} can only classify up to 90\% as accurately as their
`by eye' GZ1 training set. These constraints will have to be
taken into account for any studies that attempt to utilize morphology
in Photo-Z calculations.


\acknowledgements

The Photo-Z code used to generate the results from this paper are available on the
NASA Ames Dashlink web site https://dashlink.arc.nasa.gov/algorithm/stablegp
and is described in \cite{foster09}.

Thanks to Jim Gray, Ani Thakar, Maria SanSebastien, and Alex Szalay
for their help with the SDSS casjobs server and Jeffrey Scargle for reading
an early draft. Thanks goes to the Galaxy Group
in the Astronomy Department at Uppsala University in Sweden for their generous
hospitality where part of this work was discussed and completed.  We acknowledge
funding received from the NASA Applied Information Systems Research Program and
from the NASA Ames Research Center Director's Discretionary Fund.

This publication has been made possible by the participation of
more than 160 000 volunteers in the GZ project. Their contributions
are individually acknowledged at\newline http://www.galaxyzoo.org/
Volunteers.aspx.  Funding for the SDSS has been provided by
the Alfred P. Sloan Foundation, the Participating Institutions, the National
Aeronautics and Space Administration, the National Science Foundation,
the U.S. Department of Energy, the Japanese Monbukagakusho, and the Max
Planck Society. The SDSS Web site is http://www.sdss.org/.

The SDSS is managed by the Astrophysical Research Consortium for
the Participating Institutions. The Participating Institutions are The
University of Chicago, Fermilab, the Institute for Advanced Study, the
Japan Participation Group, The Johns Hopkins University, Los Alamos National
Laboratory, the Max-Planck-Institute for Astronomy, the
Max-Planck-Institute for Astrophysics, New Mexico State University,
University of Pittsburgh, Princeton University, the United States Naval
Observatory, and the University of Washington.

This research has made use of NASA's Astrophysics Data System Bibliographic
Services.

This research has also utilized the viewpoints \citep{GLW2010} software package.


\end{document}